# Probing the metal-insulator transition of NdNiO$_3$ by electrostatic doping

**Junwoo Son[1], Bharat Jalan[1], Adam P. Kajdos[1], Leon Balents[2], S. James Allen[2] and Susanne Stemmer[1]**

[1] Materials Department, University of California, Santa Barbara, CA 93106-5050, U.S.A.

[2] Department of Physics, University of California, Santa Barbara, CA 93106-9530, U.S.A.

**Abstract**

Modulation of the charge carrier density in a Mott material by remote doping from a highly doped conventional band insulator is proposed to test theoretical predictions of band filling control of the Mott metal-insulator transition without introducing lattice distortions or disorder, as is the case for chemical doping. The approach is experimentally tested using ultrathin (2.5 nm) $NdNiO_3$ films that are epitaxially grown on La-doped $SrTiO_3$ films. We show that remote doping systematically changes the charge carrier density in the $NdNiO_3$ film and causes a moderate shift in the metal-insulator transition temperature. These results are discussed in the context of theoretical models of this class of materials exhibiting a metal-insulator transition.

Mott insulators defy simple models that predict metallic transport. An early triumph of theory was to recognize that electron-electron repulsion can localize the conduction electrons, causing an insulator to emerge [1]. However, the connection between idealized models and real materials is complicated by structural changes, charge and magnetic ordering that often accompany the insulating state. A key probe of the Mott insulating state is to add or remove charge, which should cause a transition to the metallic state [2]. In bulk materials charge modulation is achieved via chemical substitution ("doping"), which, however, also introduces atomic disorder and lattice distortions, making it difficult to isolate the purely electronic contributions. Electrostatic doping avoids these complications, but is challenging because of the large charge carrier concentrations involved - typical Mott materials are characterized by a half-filled $d$-band, corresponding to ~ one unpaired electron per unit cell [3].

Rare earth nickelates ($R$NiO$_3$, where $R$ is a trivalent rare earth ion but not La) exhibit a first order phase transition to an insulating state upon cooling. The transition temperature ($T_{MIT}$) can be strongly modified by chemical doping with heterovalent ions on the $R$ site [4-6]. Initial reports indicated that divalent ions ("hole doping") were more effective in shifting $T_{MIT}$ to lower temperatures than quatrovalent ions ("electron doping") [5], but it was later shown that certain donor dopants also had a large effect on $T_{MIT}$ [6]. Because chemical doping also affects the Ni-O bond angles and lengths [6], an attempt to correct for the structural distortions in ref. [4] led to estimates that a 1% change in carrier concentration suppressed $T_{MIT}$ by ~ 25 to 50 °K for electron and hole doping, respectively. Separating the influence of structural distortions from band filling is particular important for the nickelates because the metal-insulator transition temperature ($T_{MIT}$) is also a strong function of the rare earth ionic radius and of applied pressure [7-9], which affect the octahedral distortions, the hybridization between O $2p$ and Ni $d$-states and the bandwidth. Isotope studies point to a strong coupling to the lattice [10]. Although the $R$NiO$_3$'s are usually classified as charge-transfer Mott insulators [8,11], more recent studies consider them to be band insulators, where charge ordering leads to a band gap between the empty $e_g$ band of a weakly magnetic Ni and the fully occupied spin-up band of the strongly magnetic Ni [12,13].

Recently, electrolyte gating, which can induce carrier densities of the order of $10^{14}$ cm$^{-2}$, has been used to induce shifts in $T_{MIT}$ of NdNiO$_3$ films [14,15]. However, electrolytes are also known to chemically alter the underlying channel [16,17], even in a reversible fashion. In ref. [14] it was reported that the observed shift in $T_{MIT}$ was a function of the film thickness, which is not consistent with a purely electrostatic effect that should produce a metallic channel at the interface that shunts the bulk of the film.

Modulation doping using a heterointerface is an attractive alternative approach for electrostatic control of large carrier densities of Mott materials and to probe the sensitivity of the transition to band filling without introducing disorder or lattice distortions [18-20]. For a 1 eV band offset, which is realistic for $R$NiO$_3$/SrTiO$_3$ interfaces [21,22], and no interface traps, estimates [23] show that the *interfacial* carrier concentration in a $R$NiO$_3$ film can be modulated greater than 20 % if interfaced with a SrTiO$_3$ film that is doped with $10^{21}$ cm$^{-3}$ electrons, far exceeding the few percent required for inducing substantial changes in $T_{MIT}$ that the bulk doping studies indicate. Although the transferred charge drops rapidly away from the interface [23], the average doping in the 2.5 nm film is estimated to be ~ 6 % and thus substantial in terms of modulation of $T_{MIT}$, by the bulk doping experiments. Carrier concentrations as high as $10^{21}$ cm$^{-3}$ are easily achievable in high-quality SrTiO$_3$ [24]. Although some charge will likely be lost to interface traps, $R$NiO$_3$/SrTiO$_3$ interfaces should be an ideal test structure to evaluate the influence of carrier modulation on the transition. In this Letter, we demonstrate substantial modulation of the carrier concentration in NdNiO$_3$, using charge transfer from a SrTiO$_3$ film and study its influence on the metal-insulator transition.

LaAlO$_3$ was chosen as a substrate because it imposes a compressive strain, which stabilizes the metallic phase of ultrathin NdNiO$_3$ films [25,26]. 5-nm-thick SrTiO$_3$ films doped with different amounts of La were grown on 5 nm undoped SrTiO$_3$ buffer layers on (001) LaAlO$_3$ by molecular beam epitaxy (MBE), described elsewhere [27]. The La-dopant concentrations ($N_{La}$) were 0, $10^{19}$, $10^{20}$ and $10^{21}$ cm$^{-3}$, respectively. The total thickness of the SrTiO$_3$ was limited to 10 nm to ensure that they were coherently strained to the substrate. A 5-nm-thick La-doped SrTiO$_3$ should not contribute to in-plane transport, because for carrier concentrations up to $10^{21}$ cm$^{-3}$ the depletion width $w$ is greater than the doped film thickness:

$$ w = \sqrt{\frac{2\varepsilon_{STO}\phi_{STO}}{eN_{La}}}, \tag{1} $$

where $\varepsilon_{STO}$ is the dielectric constant of SrTiO$_3$ ($\sim 300\varepsilon_0$ at room temperature), $\phi_{STO}$ is the band-bending ($\sim 1$ V) and $e$ is the elemental charge. Confirmation of the fully depleted nature comes from electrical measurements, described below. Epitaxial NdNiO$_3$ thin films were then grown using RF magnetron sputtering with a total gas pressure of 200 mTorr and a substrate temperature of 700 °C. The Ar/O$_2$ sputter gas ratio was 75/25, and the growth rate was $\sim 12$ nm/hour. Samples were annealed for 30 min under flowing oxygen at 600 °C after growth. All films were epitaxial, with atomically flat surfaces, as confirmed by x-ray diffraction and atomic force microscopy [23]. Measurements of sheet resistance and Hall coefficient were performed in Van der Pauw and Hall bar geometries, respectively. To obtain the Hall coefficient, $R_H$, the Hall resistance $r_H$ needed to be symmetrized to remove contributions from the magnetoresistance, i.e., $r_H = \left[ V_H \left( B \right) - V_H \left( -B \right) \right] / 2I$, where $V_H \left( B \right)$ and $V_H \left( -B \right)$ are the Hall voltages at positive and negative magnetic fields $B$, respectively, and $I$ is the current.

Due to the large carrier density and short screening length of $NdNiO_3$, ultrathin films are required to substantially modulate the charge in the film. As the sheet resistance of the thin films exceeds the Mott minimum conductivity criterion for metallic conductivity ($R_S$ < 10 kΩ/ ) localized characteristics are expected at all temperatures and are observed [25]. Figure 1 shows resistivity as a function of temperature and film thicknesses for $NdNiO_3$ films grown directly on $LaAlO_3$. 12 nm-thick $NdNiO_3$ films show a first order transition with hysteresis at $T_{MIT}$ ~ 100 K, similar to the literature [14,26]. Thin $NdNiO_3$ films show higher resistivity at room temperature and the width of the hysteresis loop becomes narrower and eventually vanishes for the 3 nm thick $NdNiO_3$ film. However, even the localized films show a sharp rise (two orders of magnitude) in resistance at $T_{MIT}$ that is not characteristic for a disorder-induced transition. The robustness of the transition to a highly resistive state even in ultrathin films makes them thus suitable to study the influence of carrier modulation [14,15].

Figure 2(a) shows the temperature-dependent sheet resistance of 2.5 nm-thick $NdNiO_3$ films grown on $La:SrTiO_3$ with the different La doping concentrations. The resistance between Hall bar structures after etching the $NdNiO_3$ was measured to confirm that no conduction occurred through the La:SrTiO₃ layers. The NdNiO₃ sheet resistance systematically decreases as the doping in the SrTiO₃ is increased, consistent with electron transfer from the La:SrTiO₃ and the schematic band alignment shown in Fig. 2(b). The NdNiO₃ films on La:SrTiO₃/LaAlO₃ with the highest La dopant concentration show a decrease in resistance on cooling above $T_{MIT}$. This is consistent with the sheet resistance of these films being significantly below 10 kΩ/  . To quantify the electron transfer across the interface, Fig. 3 shows the room-temperature Hall resistance as a function of magnetic field for the three NdNiO₃ films on differently doped La:SrTiO₃. The Hall coefficient is positive, i.e. the Hall resistivity increases linearly with magnetic fields up to 7 T for all films. The $R$NiO₃'s show semimetallic characteristics above $T_{MIT}$ [8,12]. The positive Hall coefficient is due to a large hole Fermi surface around the R point [28,29]. The Hall coefficient for relatively simple, semiconductor like material, with free electron and hole band dispersion, is given by:

$$R_H = \frac{-n\mu_n^2 + p\mu_p^2}{e\left(n\mu_n + p\mu_p\right)^2},$$  (2)

where $n$ and $p$ are the electron and hole concentrations, respectively, and $\mu_i$ their mobilities. Despite the limitations of Eq. (2), the observed positive Hall coefficient for the undoped material suggests that $p\mu_p^2 > n\mu_n^2$ and its Hall coefficient is consistent with transport dominated by holes at a carrier concentration of approximately 1 carrier per Ni. The Hall coefficient changes by about a factor of four (from 4.2×10⁻⁴ to 1.1×10⁻⁴ cm³/C) between the films on undoped SrTiO₃ and those on SrTiO₃ doped with 10²¹ cm⁻³ La. This indicates a large compensation of the hole transport by electrons. Thus both resistivity and Hall measurements demonstrate highly effective remote doping of an ultrathin, rare earth nickelate by interfacing it with highly doped $SrTiO_3$.

The results provide insights into the metal-insulator transition of the rare earth nickelates. In particular, Figure 2 shows that: (1) $NdNiO_3$ films transition to a highly resistive state below ~ 100 K even when heavily doped, and (2) $T_{MIT}$ is only weakly dependent on doping. A plot of $d\left(\ln R\right)/dT$ shows a ~ 20-30 K shift for the film on $SrTiO_3$ doped with $10^{21}$ cm$^{-3}$ La relative to films on undoped $SrTiO_3$ [23], *despite* the large change in Hall coefficient. This can be compared with the much larger effects of comparable chemical doping (a ~120 K shift for 10% doping [4]) but is similar to the shifts obtained with electrolyte gating [14,15]. The first observation is the most challenging to understand. Theoretically, a defect-free material with a finite deviation from an integral number of electrons per unit cell must be conducting at zero temperature, owing to the presence of partially filled bands. A possible explanation of the robust insulating state in the nickelates is that the stoichiometric charge and spin order of undoped $NdNiO_3$ [11-13,29,30] evolves into an *incommensurate* ordered state in concert with doping. If the wavevector of the charge order tracks the change in carrier density, an insulating ground state can be achieved. For example, bulk single-layer nickelates $La_{2-x}Sr_xNiO_4$ form insulating "stripe" states with an incommensurability which is proportional to $x$ [31]. The robustness of the $T_{MIT}$ with doping is also striking. This provides further evidence that the metal-insulator transition in the rare earth nickelates is closely tied to the spontaneous symmetry breaking and charge/spin ordering below $T_{MIT}$ [30]. In contrast, a Mott material in the strong $U$ limit described using only single-site physics would respond to substantial doping by exhibiting a crossover, as distinct from a phase transition, to a metallic phase. Finally, the modulation doping approach presented here is entirely general and could be applied to elucidate the nature of metal-insulator transitions in other materials systems.

This work was supported by a MURI program of the Army Research Office (grant # W911-NF-09-1-0398). We thank SungBin Lee and Andy Millis for many useful discussions.

concentration and plots of d(lnR)/dT as a function of temperature for NdNiO$_3$ films on La-doped SrTiO$_3$ with different carrier concentrations.

**Figure Captions**

**Figure 1:**

Resistivity of epitaxial $NdNiO_3$ thin films grown on $LaAlO_3$ as a function of temperature and $NdNiO_3$ film thickness.

**Figure 2:**

(a) Sheet resistance of epitaxial, 2.5-nm thick $NdNiO_3$ thin films on La-doped $SrTiO_3$ as a function of La-dopant concentration in the $SrTiO_3$ (see legend). (b) Schematic of the expected band lineup at the interface between metallic $NdNiO_3$ and $La:SrTiO_3$. The dashed lines indicates the position of a half filled d-band, as would be expected for an ideal Mott insulator.

**Figure 3:**

Symmetrized Hall resistances for 2.5-nm thick $NdNiO_3$ thin films on La-doped $SrTiO_3$ as a function of La dopant concentration in the $SrTiO_3$. The slope (lines) gives the Hall coefficient, which is positive and decreases with increasing La dopant concentration.

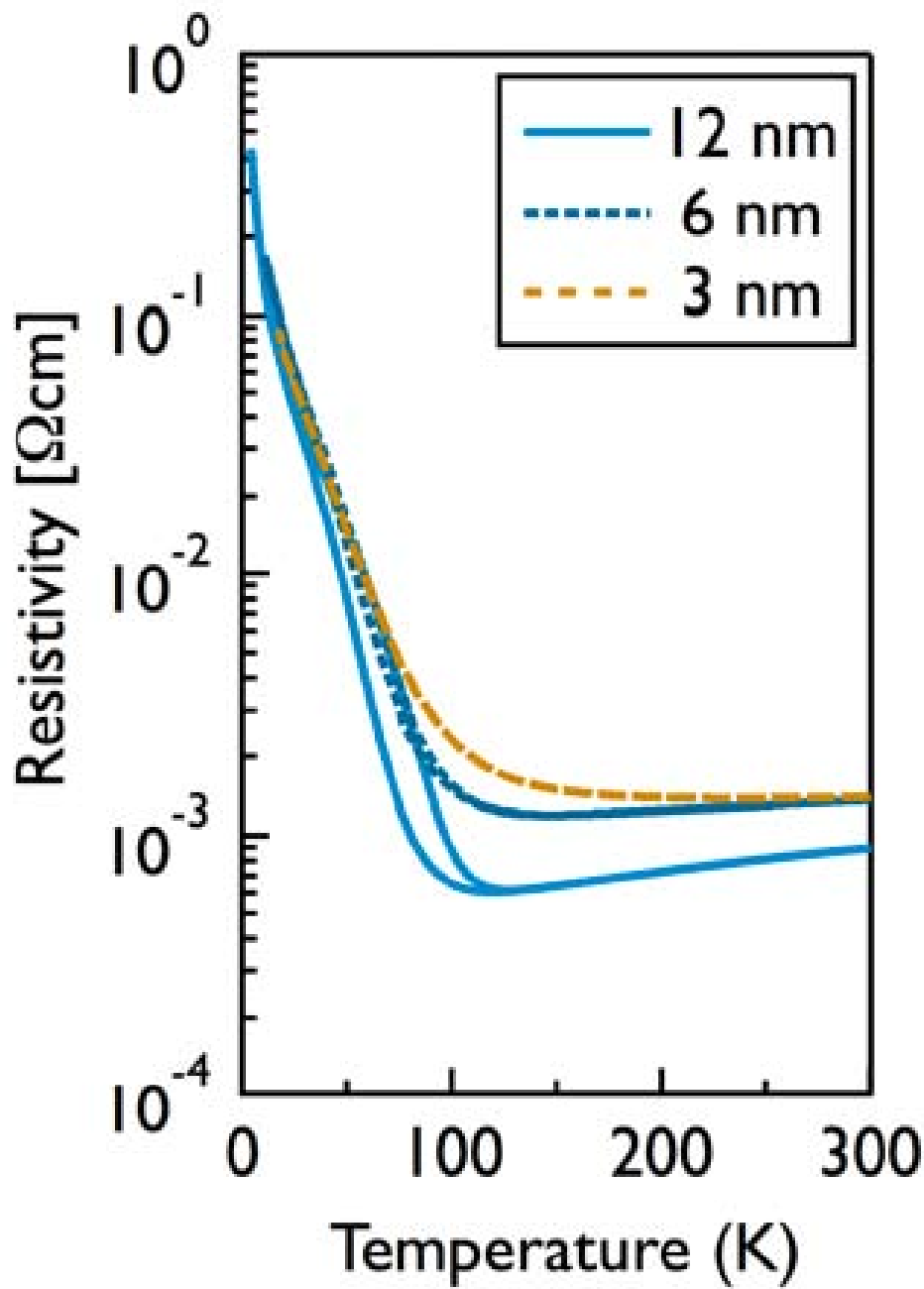

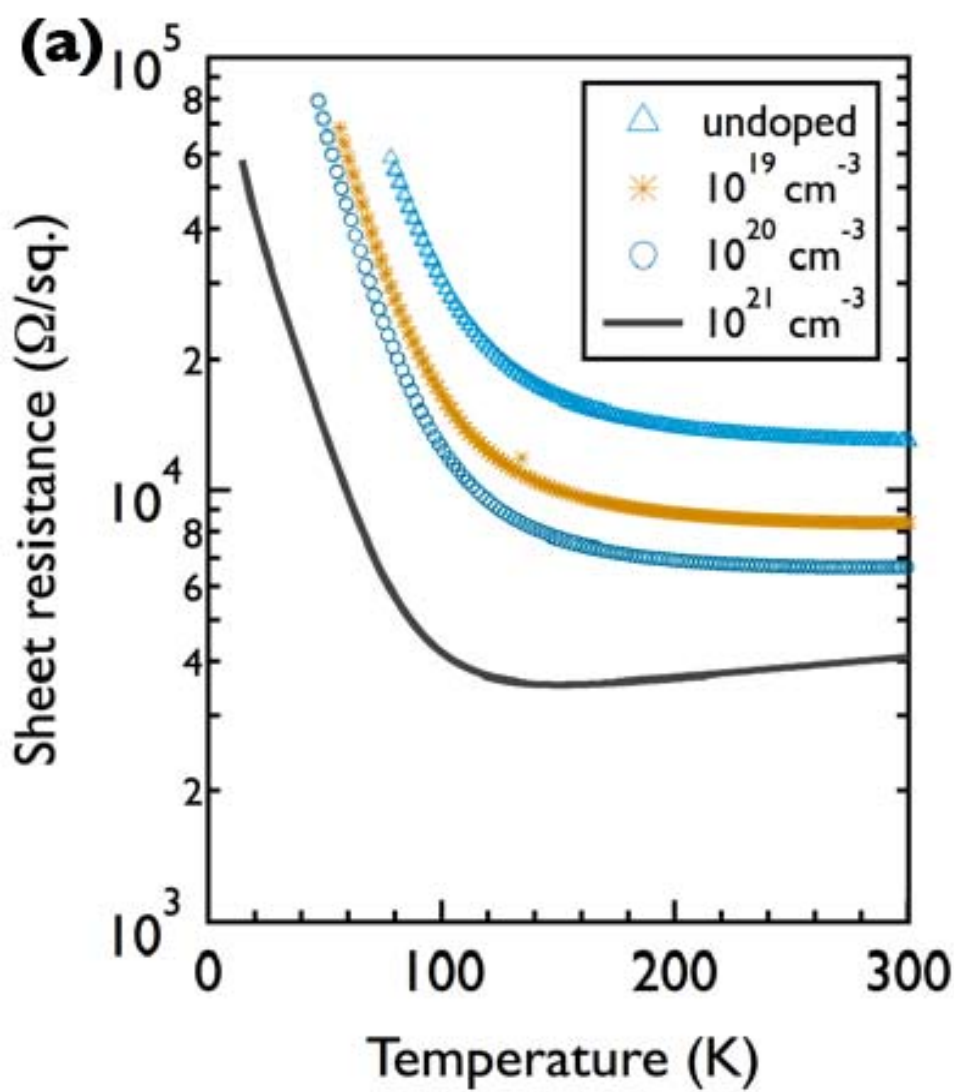

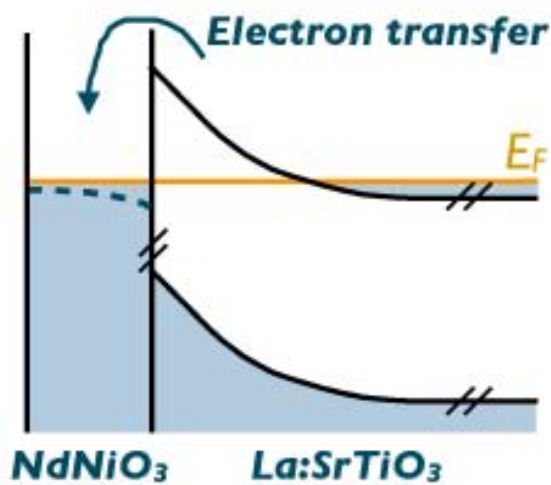

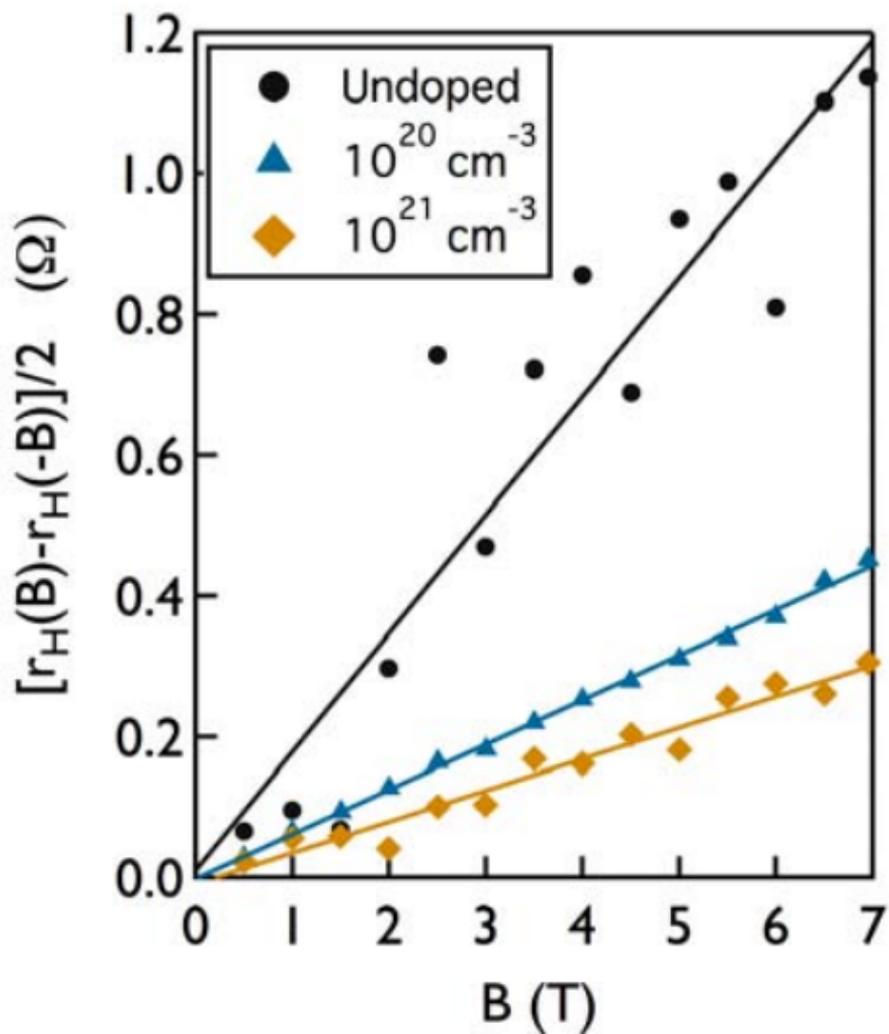

# Supplementary information

**Junwoo Son[1], Bharat Jalan[1], Adam Kajdos[1], Leon Balents[2], S. James Allen[2], and Susanne Stemmer[1]**

[1] Materials Department, University of California, Santa Barbara, CA 93106, U.S.A.
[2] Department of Physics, University of California, Santa Barbara, CA 93106-9530, U.S.A.

**Structural characterization of epitaxial NdNiO$_3$ films**

Figure S1 shows a high-resolution x-ray diffraction (XRD) radial scan through the 002 reflections of an epitaxial NdNiO$_3$ films on LaAlO$_3$ [1]. Thickness fringes indicate smooth films, and are used to estimate thickness of the films. Figure S2 shows atomic force microscope (AFM) images (3 × 3 μm$^2$ scan area) of three NdNiO$_3$ films with different thickness on LaAlO$_3$.

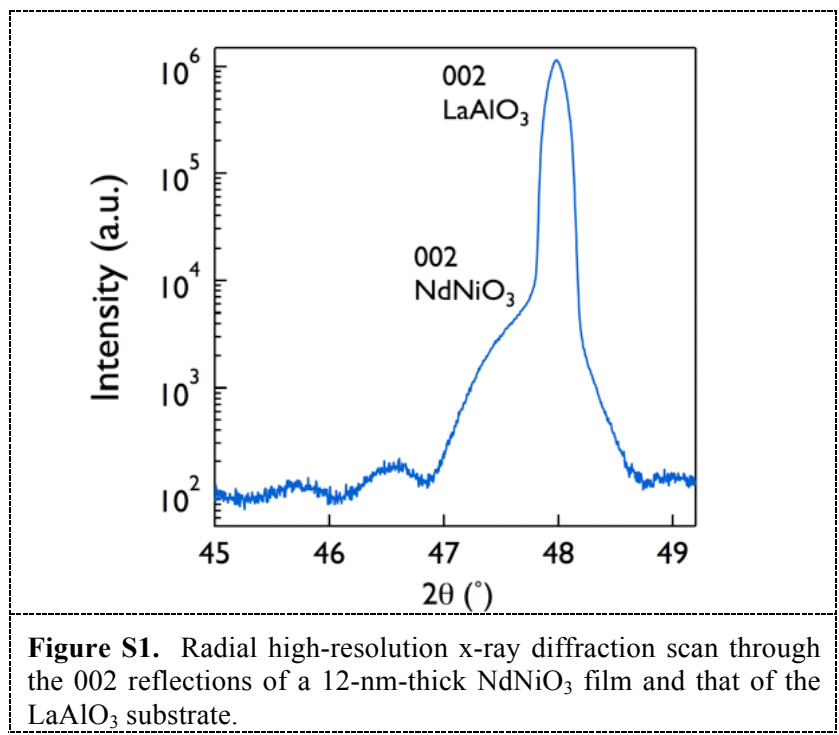

**Figure S1.** Radial high-resolution x-ray diffraction scan through the 002 reflections of a 12-nm-thick NdNiO$_3$ film and that of the LaAlO$_3$ substrate.

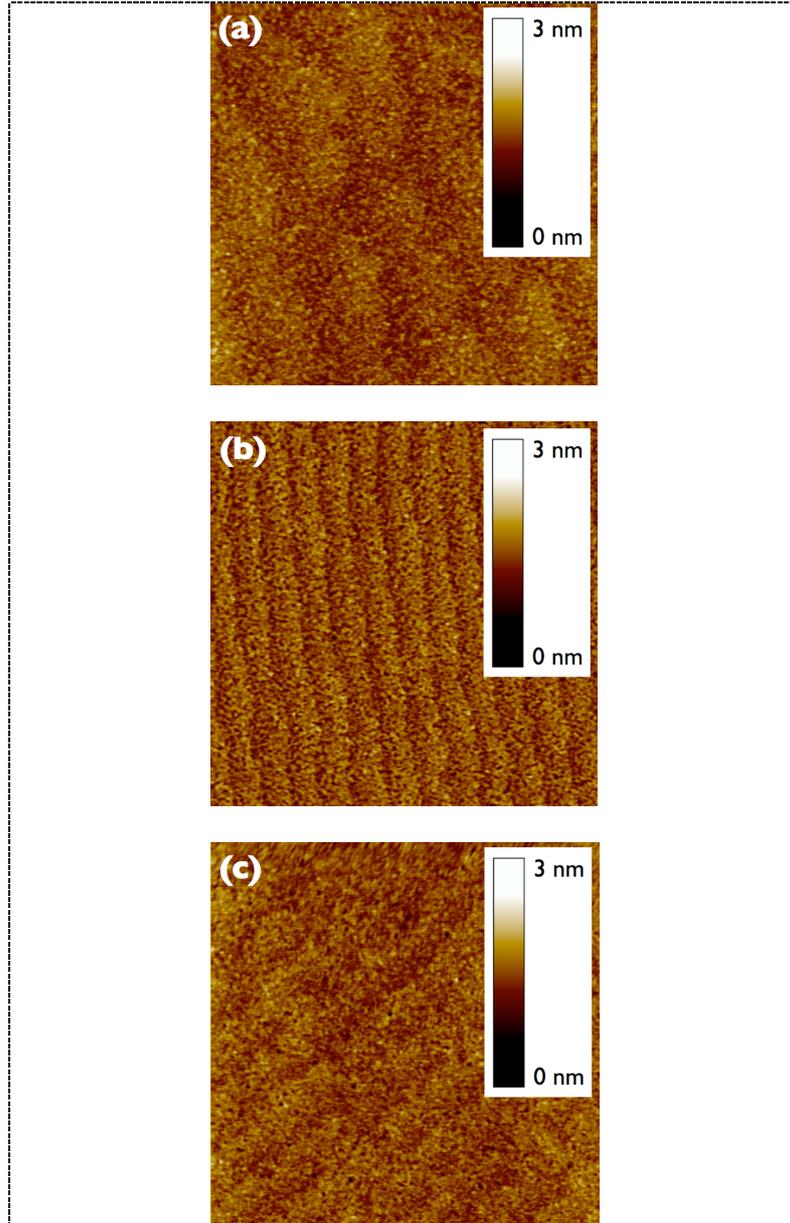

**Figure S2.** Atomic force microscopy images of the surface of (a) 3 nm, (b) 6 nm and (c) 12 nm-thick NdNiO$_3$ films on LaAlO$_3$. The root-mean-square (RMS) roughnesses are 0.161, 0.190 and 0.171 nm, respectively.

**Estimate of the depletion width of La-doped SrTiO$_3$ films**

Fig. S3 shows the SrTiO$_3$ depletion width as a function of La doping concentration ($N_{La}$) from the following relationship, as estimated from Eq. (1) in the text. The dielectric constant of SrTiO$_3$ was taken to be $300\varepsilon_0$ (room temperature value) and the band bending ($\phi_{STO}$) in the SrTiO$_3$ ~ 1 eV. Figure S3 shows that the depletion width is greater than 5 nm for doping concentrations of $10^{21}$ cm$^{-3}$ and below.

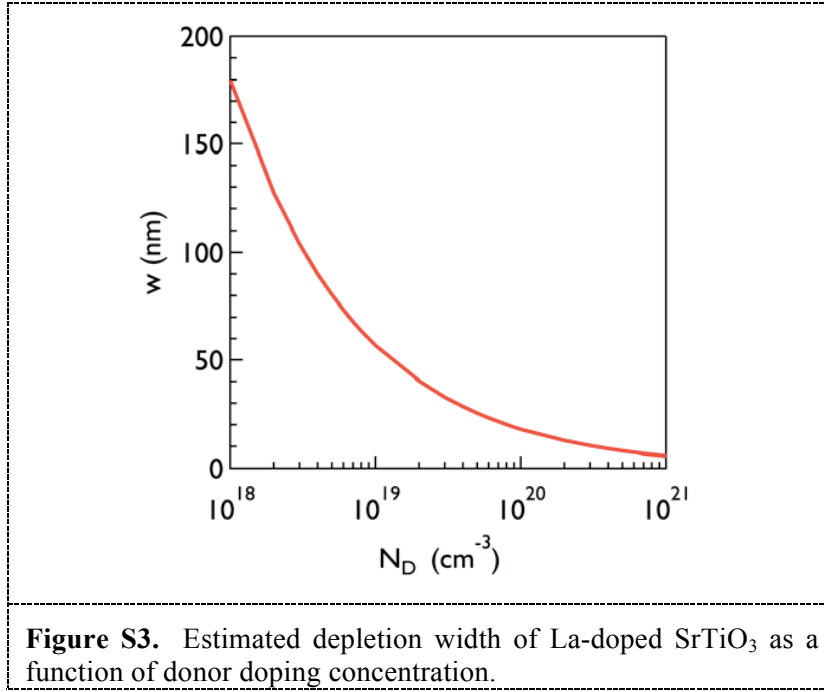

**Figure S3.** Estimated depletion width of La-doped SrTiO₃ as a function of donor doping concentration.

**Estimate of the electrostatic doping of ultrathin NdNiO₃ films**

The electrostatic doping of a 2.5 nm-thick NdNiO₃ film by interfacing it with a 5 nm-thick La-doped SrTiO₃ film can be estimated as follows. Assuming band offset ($V_B$) of 1 eV, the density of carriers transferred into the NdNiO₃ exponentially decreases with distance from the interface, which can be described using a Thomas-Fermi screening length, $k_{TF}^{-1}$:

$$k_{TF} = \sqrt{\frac{e(dn/d\mu)}{\varepsilon_{RNO}}} \; , \qquad (S.1)$$

where $e$ is the electron charge, $\varepsilon_{RNO}$ is the dielectric permittivity of the nickelate (taken to be $30\varepsilon_0$) and $dn/d\mu$ is the density of states provided by the *two* hybridized $e_g$ orbitals per nickel (per unit cell). Assuming a band width for the oxygen-hybridized $e_g$ states of order of 6 eV, the density of states is estimated to be $5.2\times10^{27}$ m⁻³V⁻¹, and the Thomas-Fermi screening length is about 0.6 nm. The transferred electrons will equal in number those depleted in the La:SrTiO₃, $n_{STO} = n_{NNO}$ with:

$$n_{NNO} = \sqrt{\left(\frac{dn}{d\mu}\right)\frac{\varepsilon_{NNO}}{e}\phi_{NNO}}\left[1-\exp\left(-k_{TF}t\right)\right], \qquad (S.2)$$

$$n_{STO} = \sqrt{\frac{2\varepsilon_{STO}N_{La}\phi_{STO}}{e}} \; , \qquad (S.3)$$

where $dn/d\mu$ is the density of states, $t$ the film thickness, $\phi_{NNO}$ and $\phi_{STO}$ the band bending potential in the NdNiO$_3$ and the SrTiO$_3$ at the interface, respectively, $e$ the elementary charge, $N_{La}$ is the La dopant concentration and $\varepsilon_{NNO}$ and $\varepsilon_{STO}$ are the dielectric constants of the NdNiO$_3$ and the SrTiO$_3$, respectively, taken as $30\,\varepsilon_0$ and $300\,\varepsilon_0$. Moreover, the transferred charge is constrained by potential continuity:

$$\phi_{NNO} + \phi_{STO} - V_B = 0 \; . \tag{S.4}$$

Figure S4 shows the estimated spatial distribution of transferred carriers from a SrTiO$_3$ film doped with $10^{21}$ cm$^{-3}$ La into a 2.5-nm-thick NdNiO$_3$ film. As expected, the density of carriers is greatest at the interface and exponentially decays with distance from the interface. The average fractional doping of the NdNiO$_3$ film is estimated to be 6 %.

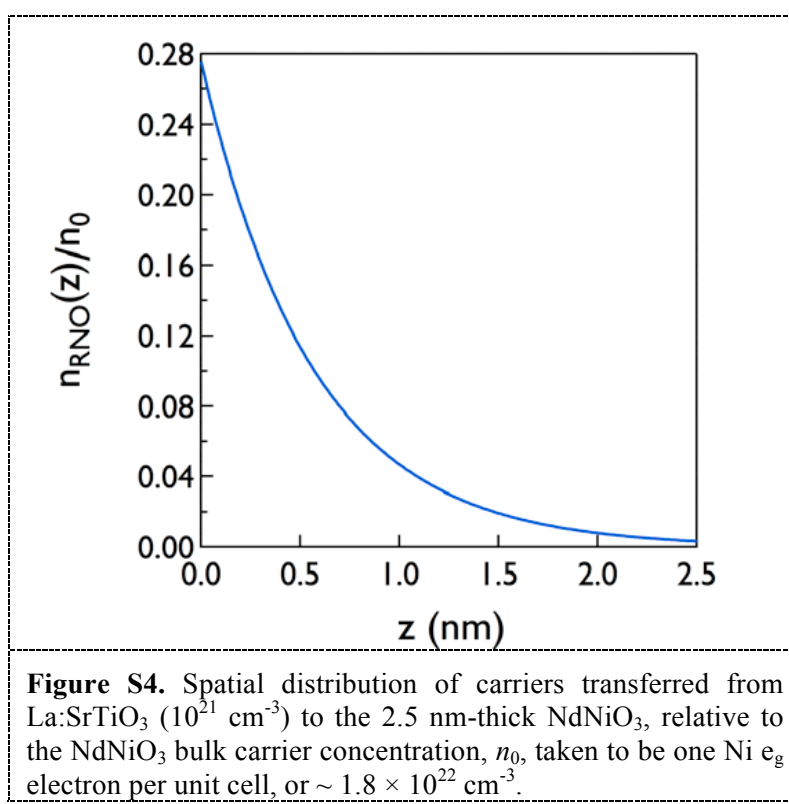

**Figure S4.** Spatial distribution of carriers transferred from La:SrTiO$_3$ ($10^{21}$ cm$^{-3}$) to the 2.5 nm-thick NdNiO$_3$, relative to the NdNiO$_3$ bulk carrier concentration, $n_0$, taken to be one Ni e$_g$ electron per unit cell, or $\sim 1.8 \times 10^{22}$ cm$^{-3}$.

**Estimate of the transition temperature from $d(\ln R_S)/dT$ plots**

Figure S5 shows plots of $d(\ln R_S)/dT$ as a function of temperature $T$, where $R_S$ is the sheet resistance of epitaxial, 2.5-nm thick NdNiO$_3$ thin films on La-doped SrTiO$_3$ with different La-dopant concentrations (see legend). The plots can serve to identify any differences in the transition temperature to the highly resistive state, $T_{MIT}$. There is almost no difference in $T_{MIT}$ for the films on the lower doped SrTiO$_3$ and a $\sim$ 20-30 K shift for the film on SrTiO$_3$ doped with $10^{21}$ cm$^{-3}$ La.

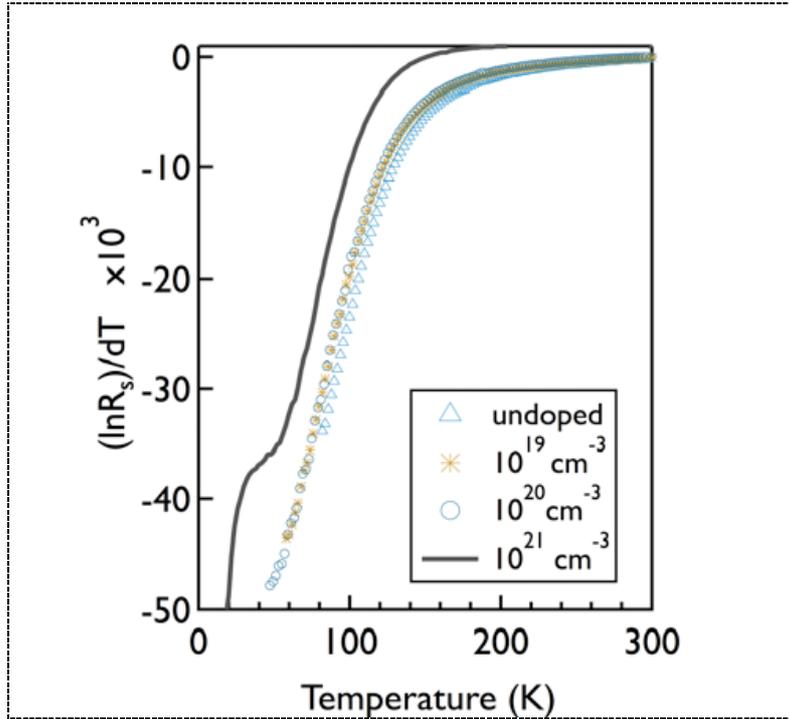

**Figure S5.** $d\left(\ln R_S\right)/dT$ as a function of temperature. $R_S$ is the sheet resistance of epitaxial, 2.5-nm thick $NdNiO_3$ thin films on La-doped $SrTiO_3$ with different La-dopant concentrations (see legend).

## References

[1] We use pseudo-cubic notation throughout.